\def\sect{\frenchspacing Section }
\def\fig{\frenchspacing Fig.~}

\documentclass[usenatbib]{mn2e} 
\usepackage{amsmath}
\usepackage{epsfig}
\usepackage{amssymb}

\title[Constraining the topology of the Universe]{Constraining the topology of the Universe using the polarised CMB maps}

\author[P.~Bielewicz, A.~J.~Banday, K.~M.~G\'orski]{P.~Bielewicz$^{1,2}$ \thanks{E-mail:
    Pawel.Bielewicz@irap.omp.eu},A.~J.~Banday$^{1,2}$,K.~M.~G\'orski$^{3,4}$\\
  $^1$ Universit\'e de Toulouse, UPS-OMP, Institut de Recherche en
  Astrophysique et Plan\'etologie, Toulouse, France \\ $^2$ CNRS, IRAP, 9 Av.~colonel
  Roche, BP 44346, F 31028 Toulouse cedex 4, France \\  $^3$ Jet
  Propulsion Laboratory, M/S 169/327, 4800 Oak Grove Drive, Pasadena
  CA 91109 \\ $^4$ Warsaw University Observatory, Aleje Ujazdowskie 4,
  00-478 Warszawa, Poland}

\begin{document}

\maketitle

\begin{abstract}
 We study the possibility for constraining the topology of the Universe by
 means of the matched circles statistic applied to polarised cosmic
 microwave background (CMB) anisotropy maps. The advantages
 of using the CMB polarisation maps in studies of the topology over
 simply analysing the temperature data as has been done to-date
 are clearly demonstrated. We
 test our algorithm to search for pairs of matched circles on simulated CMB maps for a universe with
 the topology of 3-torus. It is found that the noise levels of both Planck and next 
 generation CMB experiments data are no longer prohibitive
 and should be low enough to enable the use of the
 polarisation maps for such studies. For such
 experiments the minimum radius of the back-to-back matched circles which can be
 detected are determined. We also showed that the polarisation generated after
 reionisation does not have an impact on detectability of
 the matched circles.
\end{abstract}


\section{Introduction}
According to General Relativity the local properties of spacetime
geometry are described by the Einstein gravitational field
equations. However, they do not specify the global spatial geometry of
the Universe, i.e.\ its topology. This can only be constrained by
observations. The question of a possible multiply connected topology
of our cosmic space was already raised by \citet{schwarzschild:1900}
before the formulation of General Relativity as well by \citet{de
  Sitter:1917} and \citet{friedmann:1924} immediately 
following the formulation of the theory. 
However, because of a lack
of observational data probing the largest scales of the Universe,
the standard cosmological model was tacitly assumed to comprise
a universe possessing a simply-connected topology. One of the first
attempts to
constraining the Universe's topology was undertaken utilising the distribution of quasars
by \citet{fang:1985}. However, the first measurements of the CMB
anisotropy led to a 
significant improvement in constraints
\citep{stevens:1993,oliveira-costa:1995,oliveira-costa:1996}. 
Such anisotropy maps survey scales comparable to the horizon of the
observable Universe and therefore provide the best chance for detection of
the signatures of multi-connectedness. 

Moreover, in the last decade the study of topology has also drawn
considerable attention due to 
various anomalies observed on the largest angular scales
in the \emph{WMAP} data. Some of these anomalies, such as
the suppression of the quadrupole moment \citep{bennett:2003} and an alignment between the
preferred axes of the quadrupole and the octopole \citep{de
  Oliveira-Costa:2004, copi:2004, bielewicz:2005}, may be explained
as effects caused by a multi-connected topology. 

Current constraints on topology were established by searching for
the signatures of multi-connectedness in CMB sky
maps either in the spherical harmonic domain \citep{de
  Oliveira-Costa:2004,kunz:2006,kunz:2008} or directly in pixel space \citep{cornish:2004, aurich:2005, aurich:2006,
  then:2006, key:2007,bielewicz:2011}. In the latter case, one of the most
sensitive methods is the search for matched circles in the temperature
anisotropy patterns \citep{cornish:1998}. By means of this method as
applied to the \emph{WMAP} maps, a large class of topologies has been ruled out. 
However, the method is not inherently limited to temperature anisotropy
studies. It can also be applied to the CMB polarisation data. In this work,
we investigate such an application of the method. We
test it on simulated CMB maps for a flat universe with the topology of  a
3-torus, and explicitly consider the possibility for the detection of matching circle pairs for data
with an angular resolution and noise level characteristic of the
Planck \citep{planck:2005} 
and COrE \citep{core:2011}  data. The latter will be treated as a reference
mission for
the next generation of CMB experiments.

The paper is organised as follows. In \sect\ref{sec:anisotropies}
we briefly review some basic properties of the CMB 
temperature and polarisation anisotropies. A description of the simulations
for a  universe with multi-connected topology used
to test the reliability of our codes is presented in
\sect\ref{sec:simulations}. \sect\ref{sec:statistic} is devoted to a 
description of the statistic adopted in our studies. The disadvantages of the
search of the matched circles in temperature maps are presented in
\sect\ref{sec:temp_degrade} and in \sect\ref{sec:pol_search} the
application of the method to simulated polarisation maps is shown. In the
last section we present our conclusions.

\section{CMB anisotropies} \label{sec:anisotropies}
In this section we provide a brief resume on the properties of CMB temperature and
polarisation anisotropy.

\subsection{Temperature anisotropies} \label{sec:temp_anisotropies}
The CMB is observed as blackbody radiation with temperature \mbox{$T_0=2.725 \pm
0.002\ \rm{K}$} \citep{mather:1999}. After accounting for the peculiar motion of the Earth with
respect to the CMB rest frame, the radiation is seen to exhibit tiny temperature
fluctuations of order \mbox{$\Delta T/T_0 \sim 10^{-5}$}.  In the
Newtonian gauge, the temperature fluctuations in a given direction of
the sky, $\hat{\boldsymbol{n}}$, can be written as
\begin{equation} \label{eqn:temp_anisotropy} 
\frac{\Delta T}{T_0}\left(\hat{\boldsymbol{n}}\right) = \left(\frac{\delta_\gamma}{4}+\psi\right)_{|\tau_r}+
\hat{\boldsymbol{n}}\cdot \boldsymbol{v}_{|\tau_r}+\int_0^{\tau_0} e^{-\kappa(\tau)}(\dot{\psi}+\dot{\phi}) d\tau \ ,
\end{equation}
where $\delta_\gamma$ is the fractional energy density fluctuations in the CMB, $\boldsymbol{v}$ is 
the baryon velocity, $\phi$ and $\psi$ are the two gravitational potentials in terms of which 
the metric fluctuations are given by $ds^2=a^2(\tau)\left[-(1+2\psi)d\tau^2+(1-2\phi)dx^2 \right]$, 
and the optical depth $\kappa(\tau)=\int_{\tau_r}^{\tau_0} a(\tau') \sigma_T n_e x_e d\tau'$. $\tau$ is conformal 
time with subscripts 0 and $r$ denoting the present time and the time of recombination, respectively.

The different terms in equation (\ref{eqn:temp_anisotropy}) describe different physical effects. 
The first term corresponds to photon energy
density fluctuations and gravitational potential redshift (so called the Sachs-Wolfe effect, SW), the second
to the Doppler effect caused by baryon motions at recombination and
the third is the so-called the 
integrated Sachs-Wolfe effect (ISW) caused by the variation of the gravitational potential with time. 
The equation does not include the terms coming from other secondary
effects such as the Sunayev-Zeldovich and gravitational lensing
effects. However, they are important for angular scales smaller than the
scales of interests in this paper ($\sim 20'$).

\subsection{Polarisation anisotropies} \label{sec:pol_anisotropies}
Linear polarisation in the CMB is generated by Thomson scattering of 
photons by electrons either at the moment of recombination, when photons decouple from the primordial
plasma, or during reionisation when, due to partial ionisation of the matter, free electrons reappear again in the
Universe. The former produces a pattern of polarisation anisotropy at
small angular scales while the latter 
generates structure on large scales, corresponding to the  angular scale of the Universe at the moment of scattering.
The small-scale polarisation anisotropy is 
considered to be a snapshot of the last scattering surface (LSS). 

To generate polarisation the radiation incident on the
electrons needs to have a quadrupole moment. It can be
produced either by scalar or tensor perturbations. Depending on this,
the polarisation pattern of the sky will manifest different types of
parity. The polarisation modes generated by the former will show ``electric-type'' parity
while the modes generated by the latter -- ``magnetic-type''
parity \citep{zaldarriaga:1997}. As is common, we will refer to these two types of polarisation components as
the E-modes and B-modes, respectively. We will not consider in this
paper the impact of the gravitational lensing effect on the primordial polarisation since
it is not important for the angular scales of interest here.

In the case of scalar perturbations, gradients in the velocity field, $\partial_i \boldsymbol{v}_j$, of the 
photon-baryon fluid on the LSS lead to a
quadrupole component of the temperature seen by the scatterer   
\mbox{$\Delta T(\hat{\boldsymbol{n}}') \approx \hat{\boldsymbol{n}}_i' \hat{\boldsymbol{n}}_j' \partial_i \boldsymbol{v}_j$}, 
which, through Thomson scattering, is converted into
polarisation. Then, the polarisation expressed in terms of the Stokes
parameters $Q$ and $U$ is given by \citep{seljak:1998} 
\begin{equation} \label{eqn:pol_anisotropy} 
(Q+{\rm i}U)(\hat{\boldsymbol{n}}) \propto \sigma_T \int d\hat{\boldsymbol{n}}' 
(\boldsymbol{m}\cdot\hat{\boldsymbol{n}}')^2 \Delta T(\hat{\boldsymbol{n}}') 
\propto \Delta \tau_r \boldsymbol{m}^i \boldsymbol{m}^j \partial_i \boldsymbol{v}_j|_{\tau_r} \ ,
\end{equation}
where $\sigma_T$ is the Thomson cross section, $\Delta \tau_r$ is the width of the LSS, 
$\boldsymbol{m}=\hat{\boldsymbol{e}}_1+{\rm i}\hat{\boldsymbol{e}}_2$ and $\hat{\boldsymbol{e}}_1$, $\hat{\boldsymbol{e}}_2$ are directions
perpendicular to $\hat{\boldsymbol{n}}$ that are used to define the Stokes parameters. Tensor
$\boldsymbol{m}^i \boldsymbol{m}^j$ assures that the quantities $Q\pm {\rm
  i}U$ transform under rotations of $(\hat{\boldsymbol{e}}_1, \hat{\boldsymbol{e}}_2)$ by an angle
$\gamma$ as spin 2 quantities i.e. \mbox{$(Q\pm {\rm i} U)'(\hat{\boldsymbol{n}})=e^{\mp 2 {\rm i} \gamma} (Q\pm
  {\rm i} U)(\hat{\boldsymbol{n}})$} .

The Stokes parameters are direct measurables.
However, because of their spin 2 nature, they are not
convenient for the analysis of the polarisation field, for which
the use of spin 0 maps of the E-modes and B-modes is
preferred and allows easy discrimination of the type of 
perturbations generating the polarisation
anisotropy. Such quantities can be related to the Stokes parameters by means of
spin-raising $\eth$ and spin-lowering $\bar{\eth}$ differential
operators \citep{zaldarriaga:1997}, i.e.

\begin{eqnarray}
\widetilde{E}(\hat{\boldsymbol{n}}) &=& -{1 \over 2} \left[ \bar{\eth} \bar{\eth} (Q
+ {\rm i} U)(\hat{\boldsymbol{n}}) + \eth \eth  (Q
- {\rm i} U)(\hat{\boldsymbol{n}}) \right] \ , \nonumber \\
\widetilde{B}(\hat{\boldsymbol{n}}) &=& {{\rm i} \over 2} \left[ \bar{\eth} \bar{\eth} (Q
+ {\rm i} U)(\hat{\boldsymbol{n}}) - \eth \eth  (Q
- {\rm i} U)(\hat{\boldsymbol{n}}) \right] \ . \label{eqn:qu2eb}
\end{eqnarray}
It is worth noticing that the former equation corresponds to the
divergence operator, the latter  to the rotation operator of
the spin-2 fields on the sphere. Thus, these relations can be seen as analogs of 
Gauss's law and Ampere's law\footnote{because the measured Stokes parameters
are independent of time we do not have time dependent terms in this
case} respectively, in electrodynamics, and
the E-mode and B-mode maps as analogs of the electric charges and
the current densities.  This explains why these two
components of the polarisation field are often called the ``electric'' and 
``magnetic'' modes.

To avoid the direct differentiation of the $Q \pm {\rm i} U$ maps 
decomposition can proceed in the spin 2 spherical harmonics basis
${}_{\pm 2} Y_{\ell m}$ i.e. \mbox{$(Q \pm {\rm i} U)(\hat{\boldsymbol{n}})
= \sum_{\ell,m} a_{\pm 2,\ell m}\ {}_{\pm 2}Y_{\ell m}
(\hat{\boldsymbol{n}})$}. Then, the relations (\ref{eqn:qu2eb})
expressed by the appropriate spherical harmonics coefficients take
the form

\begin{eqnarray} \label{eqn:qu2eb_sph}
a^E_{\ell m} = - { a_{2,\ell m} +a_{-2,\ell m} \over 2} \ ,\  a^B_{\ell m} = {\rm i} { a_{2,\ell m} -a_{-2,\ell m} \over 2} \ .
\end{eqnarray}
The $a^E_{\ell m}$ and $a^B_{\ell m}$ coefficients are then related to
the E-mode and B-mode maps by

\begin{eqnarray}
\widetilde{E}(\hat{\boldsymbol{n}}) &=& \sum_{\ell m} \sqrt{{(\ell+2)! \over (\ell
  -2)!}}\, a^E_{\ell m}\, Y_{\ell
  m}(\hat{\boldsymbol{n}}) \ , \nonumber \\
\widetilde{B}(\hat{\boldsymbol{n}}) &=& \sum_{\ell m} \sqrt{{(\ell+2)! \over (\ell
  -2)!}}\, a^B_{\ell m}\, Y_{\ell
  m}(\hat{\boldsymbol{n}}) \ . \label{eqn:eb_sph}
\end{eqnarray}
It is important to note that the $a^{E,B}_{\ell m}$ coefficients are not exactly equal to the
spherical harmonic decomposition coefficients of the E and B-modes
maps. The use of the $a^{E,B}_{\ell m}$ coefficients without the factor
$\sqrt{(\ell+2)! / (\ell-2)!}$ is partly a matter of convention. It is
motivated by the fact that the power spectrum computed from such
coefficients is the same on small scales as for the Stokes parameters
themselves, 
while that of the $\widetilde{E}$ and $\widetilde{B}$ maps differs by a factor of $\propto
\ell^4$. One consequence of the $\sqrt{(\ell+2)! / (\ell-2)!}$
factor is that white noise for the Stokes parameters anisotropy fields
will transform to coloured noise for the $\widetilde{E}$ and $\widetilde{B}$ maps.

\section{Simulations of the CMB polarised anisotropy maps in the multi-connected
  universe} \label{sec:simulations}
\indent To test the reliability of the codes used to search for the signature of a
multi-connected topology in the Universe, we performed simulations of
CMB skies for a flat universe with the topology of a 3-torus
\citep{riazuelo:2004}. 
The first step in this process involved the computation of 
the coefficients of the E-modes, $a_{\ell m}^E$, in an analogous
manner to the temperature anisotropy coefficients as
described in \citet{bielewicz:2011}. The only significant difference 
was the use of the response function for the E-mode
polarisation, $\Delta_\ell^{E}(k,\Delta \tau)$, 
instead of that for the temperature in their equation (1). 
The coefficients were then convolved with an appropriate 
beam profile, and subsequently  transformed
to the experimentally observed Stokes parameters using the relation (\ref{eqn:qu2eb_sph}) and inverse of
the spin 2 spherical harmonics transform described in the previous section.
Finally, homogeneous noise was added to
the $Q$ and $U$ maps at a characteristic level for a given experiment.

In the simulations, we did not take into account the B-modes for multi-connected
universes. The amplitude of the polarisation generated by gravitational waves is much smaller than
the E-mode polarisation and the noise level for the Planck data is too
high to allow the use of these maps for constraining the topology. 
We also do not consider effects that generate B-mode polarisation from the gravitational 
lensing of the E-modes. As is the case for any secondary effect, this
dilutes signatures coming from the topology. Nevertheless, it is important for
scales much smaller than those considered in this work ($\sim
20'$). However, it is important to note that in the simulated
  maps used in this paper B-modes are present due to the noise.

To study the signatures of a given topology, a CMB map
is required with resolution comparable to the angular size of the first 
significant acoustic peaks of the polarisation map i.e.~$\ell_{\rm max} \sim
400-500$. We have adopted $\ell_{\rm max} = 500$ in our
simulations. The dimension of the fundamental domain  
of the 3-torus was $L=2\, c/H_0$, which is about three times less than
both the diameter of the
observable Universe i.e.\ $\sim 6.6\, c/H_0$, and 
current lower bounds on the size of the Universe obtained from
a matched circles study of  temperature anisotropy maps
\citep{key:2007,bielewicz:2011}. Although such a small 3-torus does not
fit very well to the data, the simulations are used only to
demonstrate and compare the performance of the statistic for matched circles over
a wide range of radii. A universe where there are many pairs of matched circles of different
radii serves very well for these objectives. The values of the cosmological parameters adopted corresponded 
to the $\Lambda CDM$ model \citep[see][Table 3]{larson:2011} determined from the
7-year \emph{WMAP} results. The time needed for the generation of one such CMB map on a single
processor with clock speed 3 GHz is about 42 hours.

\section{Statistic for the matched circles} \label{sec:statistic}
If light had sufficient time to cross the fundamental cell, an observer would see multiple copies of a single 
astronomical object. To have the best chance of seeing `around the universe' we should look for multiple 
images of distant objects. Searching for multiple images of the last scattering surface
is then a powerful way to constrain
topology. Because the surface of last scattering is a sphere centred on the observer, each copy of the observer
will come with a copy of the last scattering surface, and if the copies are separated by a distance less than the 
diameter of the last scattering surface, then they will intersect along 
circles. These are visible by both copies of the observer, but from opposite sides. The two copies are really 
one observer so if space is sufficiently small, the CMB radiation from the last scattering surface will contain
a pattern of hot and cold spots that match around the circles. 

The idea of using such circles to study topology
is due to \citet{cornish:1998}. Therein, a
statistical tool was developed to detect correlated circles in all sky maps of the CMB
anisotropy -- the circle comparison statistic
\begin{equation} \label{eqn:s_statistic}
S_{p,r}^{\pm} (\alpha, \phi_\ast)=\frac{\left<2\, X_p(\pm \phi) X_r(\phi+\phi_\ast)\right>}{\left<X_p(\phi)^2+X_r(\phi)^2\right>}\ ,
\end{equation}
where $\left< \right>=\int_0^{2\pi}d\phi$ and $X_p(\pm \phi)$,
$X_r(\phi+\phi_\ast)$ are temperature (or polarisation)
fluctuations around two circles of angular radius $\alpha$  centered
at different points, $p$ and $r$, on the sky with relative phase
$\phi_\ast$. The sign $\pm$ depends 
on whether the points along both circles are ordered in a clockwise direction
(phased, sign $+$) or alternately whether along one of the circles the
points are ordered in an anti-clockwise direction (anti-phased, sign
$-$). This allows the detection of both orientable and non-orientable topologies.
For orientable topologies the matched circles have anti-phased
correlations while for non-orientable
topologies they have a mixture of anti-phased and phased correlations.
The statistic has a range over the interval $[-1,1]$. Circles that are 
perfectly correlated or anticorrelated have $S=1$ or $S=-1$, respectively, while uncorrelated circles will have a
mean value of $S=0$. To find correlated circles for each radius $\alpha$, the
maximum value $S_{\rm max}^{\pm}(\alpha) = \underset{p,r,\phi_\ast}{\rm max}
\, S_{p,r}^{\pm}(\alpha,\phi_\ast)$ is determined. In case of
anticorrelated circles the maximum value of
$-S_{p,r}^{\pm}(\alpha,\phi_\ast)$ is used. 

In the original paper by \citet{cornish:1998} the above statistic was
applied exclusively to temperature anisotropy maps. However, it can
also be applied to polarisation data and in this work, we focus on
its application to the E-mode map. In this case, the $X$ map is simply
the map of the E-mode.

In order to speed up the computations, one can use the fast Fourier transform (FFT) along the circles, 
$X_p(\phi) = \sum_m X_{p,m} \exp({\rm i}m\phi)$, and, by rewriting the statistic as 
$S_{p,r}^{+}(\alpha, \phi_\ast) = \sum_m s_m \exp(-{\rm i}m\phi_\ast)$, where 
\mbox{$s_m = 2 \sum_m X_{p,m}^{} X_{r,m}^\ast / \sum_n \left( |X_{p,n}|^2+|X_{r,n}|^2\right)$}, use the
inverse fast Fourier transform. It reduces the computational cost
$S_{p,r}^{\pm}(\alpha,\phi_\ast)$ from $\sim N$ to  $\sim N^{1/2} \log N$
operations, where $N$ is the number of pixels in the map. 

In our studies we will use the statistic as modified in
\citet{cornish:2004} to include weighting of the $m$th harmonic of
the temperature around the $p$th circle $X_{p,m}$ by the factor $|m|$
such that the number of degrees of freedom per mode
are taken into account.
\begin{equation} \label{eqn:s_statistic_fft}
S_{p,r}^{+}(\alpha, \phi_\ast)=\frac{2 \sum_m |m| X_{p,m}^{} X_{r,m}^\ast
e^{-{\rm i} m \phi_\ast}}{\sum_n |n| \left( |X_{p,n}|^2+|X_{r,n}|^2\right)}\ ,
\end{equation}
where the Fourier coefficients $X_{p,m}$ are complex conjugated in the
case of the $S_{p,r}^{-}$ statistic. Such weighting enhances the
contribution of small scale structure relative to the large scales
fluctuations. It is especially important because the large scale fluctuations,
which are dominated by the integrated Sachs-Wolfe (ISW) effect for
the temperature anisotropy and the reionisation effect in the case of
polarisation, can obscure the image of the last scattering surface
and reduce the ability to recognise possible matched patterns on it.

The general search explores a six dimensional space: the location of the
first circle centre, $(\theta_p, \phi_p)$, the location of the second circle centre, $(\theta_r, \phi_r)$, the 
angular radius of the circle $\alpha$, and the relative phase of the two circles $\phi_\ast$. The number of 
operations needed to search for circle pairs scales as $\sim N^3 \log N$; thus 
implementation of a matched circles test is computationally very
intensive. As a consequence, we restrict our analysis to a search for
pairs of circles centered around antipodal points, so-called
back-to-back circles. Then, the search space can be reduced to four dimensions
and the number of operations to $\sim N^2 \log N$. Although the back-to-back search can 
detect a large class of the topologies that we might hope to find, there remain
those topologies that predict matched circles without antipodal configurations such as the 
Hantzsche-Wendt space for flat universes or the Picard space for hyperbolic universes 
\citep{aurich:2004}.

As in \citet{bielewicz:2011}, we used the
\textsc{healpix}\footnote{http://healpix.jpl.nasa.gov} scheme with
a resolution parameter $N_{side} = 512$ to define our search grid on the
sky. To accommodate the use of the FFT approach,
$n=2^{r+1}$ points are used for each circle, where $r$ corresponds to
the resolution
parameter of the map given by $N_{side} = 2^r$. By definition this is
also the angular resolution used to step through $\phi_{\ast}$. For $\alpha$, we used steps
a little bit smaller than the characteristic scale $\theta_c$ of
coherent structures in the map, i.e.\ $2\, \theta_c / 3$. As we
discuss below, to increase the signal-to-noise ratio the polarisation maps
were smoothed with a gaussian beam profile with FWHM of 50'. 
Then, the scale of coherence is defined relative to the 
two-point correlation function as twice its angular width 
at half the maximum of the zero lag value.
For polarisation maps this is around 20'. 
Contrary to the temperature
anisotropy this scale is smaller than the
smoothing scale due to the bluer power spectrum,
dominated by small angular scales, of the polarisation
maps. The values of the E-mode map at each point along the circle were
interpolated based on values for the four 
nearest pixels with weights inversely proportional to the distances between
a given point and the pixel centers. Other methods of interpolation
can be used but we have verified that even using the value
of the nearest pixel does not change the value of the statistic significantly.
The time needed for a search of one map using this scheme 
on a single processor with clock speed 3 GHz is about 180 hours.

To draw any conclusions from an analysis based on the statistic $S_{\rm
 max}^{\pm}(\alpha)$ it is important to correctly estimate the threshold
for a statistically significant match of the circle pairs. The chance of random
matches is inversely proportional to the number of coherent structures
along the circles, therefore a false positive signal level of the statistic is
especially large for circles with smaller radii. We
used simulations of the maps with the same noise properties and
smoothing scales as the Planck data \citep{planck:2005} to establish the threshold
such that fewer than 1 in 100 simulations would yield a false event. 

It is important to note that the false detection level is smaller for
higher resolution maps. Conversely, as shown by \citet{cornish:1998},
the value of the statistic for matched circles 
\begin{equation} \label{eqn:signal2noise}
S_{\rm max} \approx {\xi^2 \over 1+\xi^2} \ ,
\end{equation}
depends on the ratio of the signal rms $\sigma_s$ to the noise rms
$\sigma_n$ ratio, denoted $\xi$ for the map, 
thus the efficiency of the statistic to detect matched circles is increased by smoothing the data. 
A smoothing scale should therefore be adopted that is a reasonable
trade-off between these two requirements. 

To eliminate the regions most contaminated by Galactic foreground
residuals, we utilise the polarisation analysis mask P06
defined by the \emph{WMAP} team \citep{page:2007}. However,
the statistic is very sensitive to the masking, particularly if a
significant fraction of one or both circle pairs 
lies in the masked region. In this case, there is a significant probability
to find a chance correlation of the temperature pattern between unmasked parts of the circles,
thus increasing the false detection level. The effect is most
pronounced for circles with the largest radii, close to 90 degrees, as well
as for very small radii. To avoid this, we restrict our analysis
to those pairs for which less than half the length of each circle
is masked. Although the statistic computed in this way is not optimal,
it is much more robust with respect to masking.

\section{Discussion of degrading effects for temperature maps} \label{sec:temp_degrade}
Hitherto, the search for matched circles was performed only for
the CMB temperature anisotropy maps. The main obstacle to the
the application of this method to polarisation data is the very high noise level
of the currently available CMB polarisation maps. This situation will
change with the release of the
Planck data. As we will show below, in this case the typical 
noise amplitude will be sufficiently low to
enable the use of the polarisation data in studies of topology. This provides
an opportunity to avoid some of the problems inherent in the analysis of
temperature maps.

\begin{figure*}

\centerline{
\epsfig{file=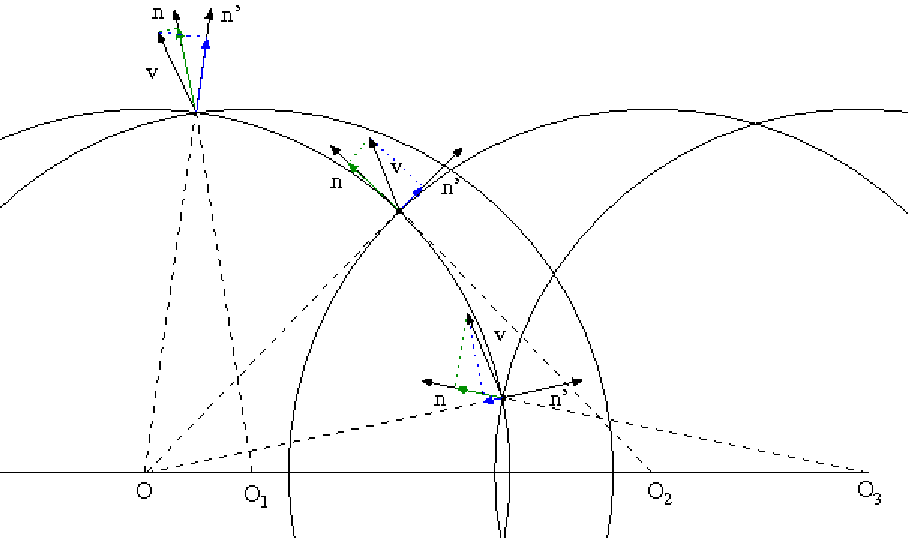,scale=0.75}
}

\caption{The baryon velocity $\boldsymbol{v}$ at the last scattering surface seen from
  two different directions $\boldsymbol{n}$, $\boldsymbol{n'}$ in the
  multi-connected universe. For the radius of the matched
  circles larger than 45$^\circ$ the projections of the baryon
  velocity on the direction of observations (dashed lines)
  for the observers $O$ and $O_1$ are strongly correlated. For the
  radius equal to 45$^\circ$ (observers $O$ and $O_2$) the velocity projections
  are independent and for the radius smaller than 45$^\circ$
  (observers $O$ and $O_3$) the projections are anti-correlated. The
  projections are denoted by blue and green arrows.} 
\label{fig:doppler_geometry}
\end{figure*}

\begin{figure*}

\centerline{
\epsfig{file=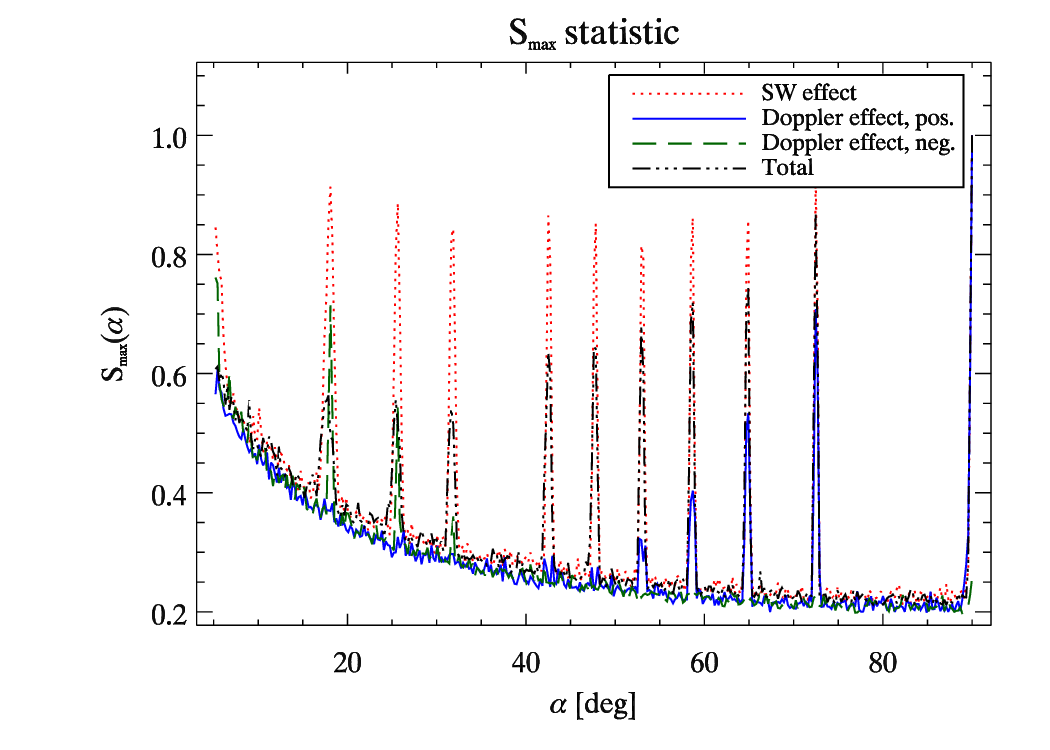,scale=0.75}
}

\caption{An example of the $S_{\rm max}^{-}$ statistic for a simulated CMB temperature anisotropy map of
  universe with the topology of a cubic 3-torus with dimensions $L = 2\ c/
  H_0$. The dotted, solid, dashed and three dot-dashed lines show the statistic for
  CMB maps of the Sachs-Wolfe (SW) effect, the positive and negative correlations
  of the Doppler effect and total anisotropy, respectively.} 
\label{fig:smax_sw_doppler}
\end{figure*}

Since the signatures of topology are imprinted on the surface of last
scattering, any effects that dilute this image will also degrade
the ability to detect such signatures by means of the matched circles
statistic. In the case of the temperature fluctuations, there are two
sources of anisotropy generated at the
LSS: the combination of the internal photon density
fluctuations and the SW effect, described by the first term of equation
(\ref{eqn:temp_anisotropy}), and the Doppler effect, described by the
second term.
In the latter case, the correlations for pairs of matched 
circles can be negative in universes with multi-connected topology. As
we can see in \fig\ref{fig:doppler_geometry}, for pairs of
back-to-back circles with 
a radius smaller than 45$^\circ$ the dot products
$\hat{\boldsymbol{n}} \cdot \boldsymbol{v}$ have opposite signs for two
nearly antipodal directions of observations. The Doppler term becomes increasingly
  anticorrelated for circles with smaller radius. One can show that on
  average in a flat universe with orientable topology the statistic (\ref{eqn:s_statistic}) for
  this term is $\left< S^{-}(\alpha) \right> = - \cos(2 \alpha)$
  \citep{riazuelo:2006}. Anticorrelation of the Doppler term
  significantly decreases correlations with the
  circle radius  for the total CMB anisotropy map. We can see
  this effect in \fig\ref{fig:smax_sw_doppler} for a simulated CMB sky
  in a universe with the topology of the 3-torus.

The ISW effect, as for any other secondary anisotropy contribution, will also
smear out signal coming from the LSS. However,
because it dominates only at large angular scales, this can be remedied
either by high-pass filtering of the map or by using the matched
circles statistic with $m$ weighting (\ref{eqn:s_statistic_fft})
to prevent the statistic from being dominated by the large angular scales.

The consequences of these degrading effects are weaker constraints on the
topology of the Universe obtained from the matched circle
statistic. As we can see in \fig\ref{fig:smax_sw_doppler}, use of the CMB map without both of the
degrading effects would allow us to impose lower bounds on the minimum radius of
the correlated circles which can be detected much below the present
constraints $\alpha_{\rm min} \approx 10^\circ$ \citep{bielewicz:2011}.

\section{Search of matched circles in polarisation maps} \label{sec:pol_search}
As already mentioned in \sect\ref{sec:pol_anisotropies}, a
polarisation map at small angular scales can be considered as a snapshot of
the LSS. Theoretically, then, the polarisation provides a better
opportunity for the detection of multi-connected topology signatures
than a temperature anisotropy map. The only serious issues preventing
its use in studies of topology are
instrumental noise and the correction of the polarised data for the
Galactic foreground. Since the polarisation anisotropy amplitude is around
ten times smaller than the amplitude for temperature, these constitute
more serious problems than for temperature anisotropy maps.

In addition, since polarisation is generated by the scattering of a quadrupolar distribution
  of radiation on free electrons at the moment of recombination or during
  reionisation, the orientation of the local quadrupole can be a
  significant problem to this analysis. The orientation of the dipole
  and quadrupole are determined
  uniquely by the initial density field. Thus, one can see from equation
  (\ref{eqn:pol_anisotropy}) that the generated polarisation depends
  on the projection of the quadrupole onto the plane perpendicular to the line
  of sight. As shown by \citet{riazuelo:2006} this projection can be
  expressed by $1-(\hat{\boldsymbol{n}} \cdot \hat{\boldsymbol{v}})^2$. This is
  analogous to the projection of the
  dipole direction on the line of sight for the Doppler term, as
  discussed in the previous section. Clearly, the contribution is
  included in our simulations, and any deleterious effect is accounted
  for in the statistical analysis. 

The other effects that might degrade the signal from the  LSS are
the polarisation generated after reionisation and the gravitational
lensing effect. The former prevails at very large angular scales where,
as for the ISW effect and temperature anisotropy data, its influence can be
eliminated by either high-pass filtering  or the use of the
statistic (\ref{eqn:s_statistic_fft}).
We will investigate this below in subsection
\ref{sec:reionisation}. The lensing effect is 
most significant at very small scales ($\sim 3'$) and is
negligible for the range of scales of interests in this paper.

We have tested the performance of the matched circles statistic for simulated
Planck \citep{planck:2005} and COrE \citep{core:2011} polarisation
maps. Although the latter has not currently been selected as a candidate
mission by ESA, we will consider it as a useful reference mission for
the next generation of observational CMB projects.

Before we present the results for simulations, it is worth considering a
rough estimation of the performance of the matched circle statistic
for different experiments. As we have already mentioned, the noise
level of the \emph{WMAP} polarisation data is too high to enable detection of
the matched circles. Even for the map assembled from the \emph{WMAP} Q, V and W-band
polarised data with inverse noise weighting we find 
$S_{\rm max} \approx 0.05$. This is considerably below the
false detection threshold of the statistic. There is a much improved
prospect for the detection of the signatures of topology for the Planck
data. As for the \emph{WMAP} data, to decrease the
noise level we consider the map obtained from  a
combination of the data with the lowest noise levels using  inverse
noise variance
weighting. Using the sensitivity and resolution of the polarised
anisotropy map derived from the 100, 143 and 217 GHz frequency maps,
i.e.~with $\Delta_N^{Q,U} \approx 59\ \mu {\rm K}\, {\rm arcmin}$ and
$\theta_{\rm FWHM} =10\ {\rm arcmin}$ \citep{planck:2005}, 
respectively, we get $S_{\rm max} \approx 0.6$. This is sufficient to enable
the detection of pairs of matched circles. Conversely, the noise level for single frequency maps
seems to be too large to allow such a test -- even the 143 GHz channel
with a sensitivity $\Delta_N^{Q,U} \approx 82\ \mu {\rm K}\,
{\rm arcmin}$ \citep{planck:2005} only renders a
value of $S_{\rm max} \approx 0.4$. The best chance of detection
for such signatures of topology arise for an experiment with
specifications similar to those of the COrE project. For a channel
with minimal contamination by the Galactic foreground, 
i.e.~105 GHz, with a polarisation sensitivity of $\Delta_N^{Q,U} \approx
4.6\ \mu {\rm K}\, {\rm arcmin}$ \citep{core:2011}, the statistic
$S_{\rm max} \approx 0.99$ achieves a value close to the maximum. 

As described in \sect\ref{sec:simulations}, we simulated the CMB polarisation
maps, in terms of the observable Stokes parameters $Q$ and $U$, for a universe with the topology 
of a 3-torus of size $L=2\, c/H_0$. The resolution of the maps and the noise rms correspond to the
Planck 100, 143 and 217 GHz frequency maps coadded with inverse noise
weighting and the COrE 105 GHz frequency map. In both cases,
the maps were convolved with a gaussian beam of FWHM=10' and
homogeneous noise with a sensitivity of $\Delta_N^{Q,U} \approx 59\
\mu {\rm K}\, {\rm arcmin}$ and $\Delta_N^{Q,U} \approx 4.6\ \mu {\rm
  K}\, {\rm arcmin}$, respectively, were added. 

Since the direct use of the Stokes parameters in the
  analysis  would require the rotation of values into a common coordinate system (defined by the pairs of points
in question) which can significantly slow down the computations, the analysis was carried 
out for E-mode maps obtained from the
$Q,U$ maps by means of the relation (\ref{eqn:qu2eb_sph}). The use of the
spin 0 maps allows us to speed up computations by the application of an FFT
along the circles, as described in \sect\ref{sec:statistic}.
The maps were also smoothed with a beam of FWHM=50' to increase the signal to
noise ratio. Because of the very blue power spectrum of
the E-mode maps, caused by the $\sqrt{(\ell+2)! /
  (\ell-2)!}$ factor (see equation (\ref{eqn:eb_sph})), in order to
remove the noise on the smallest angular scales the width of the
beam profile has to be wider than would be the case for the temperature map. The
smoothing serves also to eliminate from the analysis
multipoles of order larger than the maximum order used for the
simulated CMB maps i.e.~$\ell_{\rm  max} = 500$. 

\begin{figure*}

\centerline{
\epsfig{file=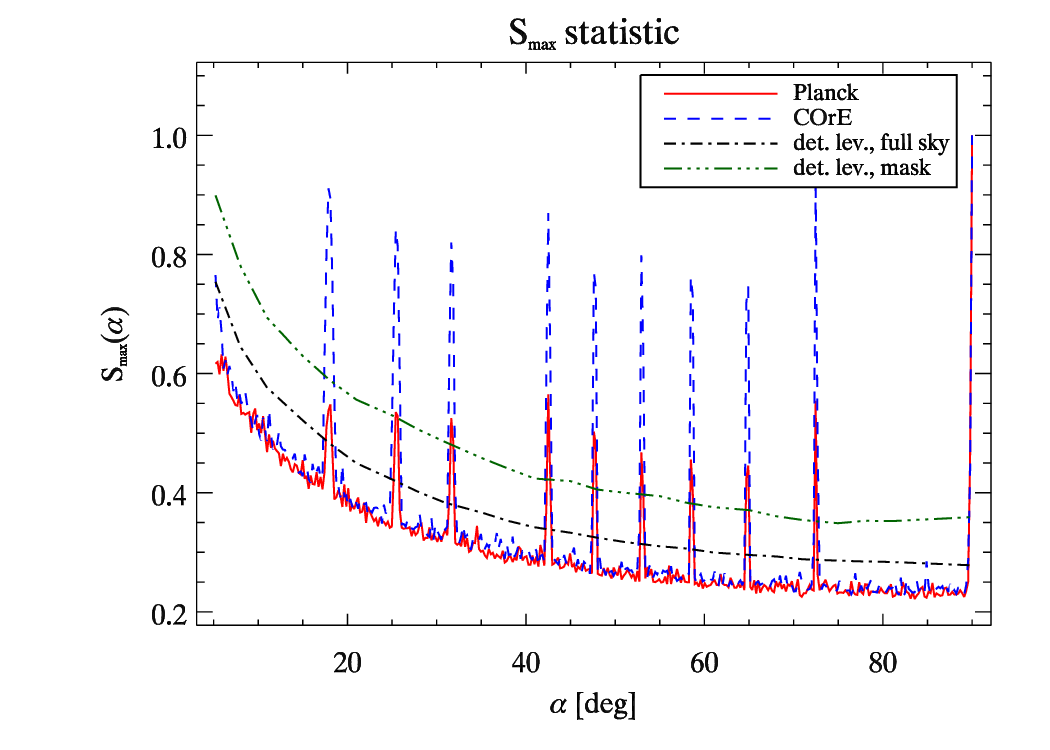,scale=0.75}
}

\caption{An example of the $S_{\rm max}^{-}$ statistic for a
    simulated polarised CMB map of the 
  universe with the topology of a cubic 3-torus of dimension $L = 2\ c/
  H_0$. The solid and dashed lines show the statistic for simulated polarisation maps with angular resolution 
and noise level corresponding to the Planck and COrE data, respectively. The dot-dashed and three 
dot-dashed lines show the false detection levels for the statistic
estimated from 100 Monte Carlo simulations 
of the Planck coadded 100, 143 and 217~GHz frequency polarisation maps for the full sky and cut sky analysis, 
respectively.}
\label{fig:smax_t222_pol}
\end{figure*}

The $S^{-}_{\rm max}$ statistic for the simulated maps is
shown in \fig\ref{fig:smax_t222_pol}. As expected the amplitudes of the
peaks do not decrease with the radius of the circles as in the case of the
temperature anisotropy maps. We see that pairs of matched circles
can be detected for the Planck coadded map. However, comparing with
\fig\ref{fig:smax_sw_doppler} one should notice that the relative amplitude of the peaks with
respect to the average correlation level for the circles with small radius is not
bigger than for the temperature map. Thus, the constraints on topology
will not be much tighter than those derived from an analysis of
temperature maps. A much better perspective arises for the COrE
maps. The signal of the multi-connected topologies is very pronounced
in this case enabling the detection of matched circles with very small
radius. It is also worth noticing that the width of the peaks is smaller
than for temperature maps. It is a consequence of domination of the
small angular scales in the power spectrum of the polarised maps
leading to more narrow peak of zero lag of the two-point correlation function.

To determine more precisely the minimum radius of the detectable
circles it is needed to estimate the false detection level of the statistic. We used
for this purpose 100 simulations of the Planck coadded map for a
universe with a simply-connected topology. The detection level was estimated for 
both the full sky $Q$ and $U$ maps and the masked maps. The former corresponds to
the case of a perfect separation of the CMB
and the Galactic foreground polarisation and yields optimal performance of the 
statistic. The latter presents a more realistic estimation taking into
account the necessity of removing those regions most 
contaminated by the Galactic foregrounds.

\begin{figure*}

\centerline{
\epsfig{file=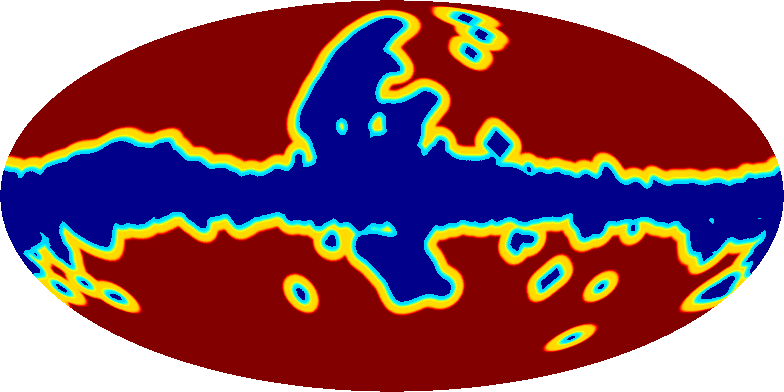,scale=0.8}
}

\caption{Masks used in the analysis. The dark blue region corresponds to
  the part of the sky unused in polarisation analysis, 
  defined by the \emph{WMAP} team in the P06 mask.
  The light blue and yellow bands correspond to
  the apodized regions of the extended mask used for the E/B decomposition and
  the mask used in the search for the matched circles,
  respectively. The red region corresponds to the unmasked and
  unsmoothed part of the sky.} 
\label{fig:masks}
\end{figure*}

We used masks based on the P06 mask derived by the 
\emph{WMAP} team \citep{page:2007}  for their analysis of polarised data.
In order to minimize ambiguous modes that arise from the decomposition of the 
Stokes parameters into the E and B-mode on incomplete sky coverage, we used the following 
procedure for computation of the E-mode maps.
The P06 mask was extended and apodized employing the method described by
\citet{kim:2011} (referred to as a 'processing mask' therein) i.e.~the mask
was widened so that after apodization with a Gaussian profile it is zero
in regions defined as such by the original mask up to a precision of
$10^{-6}$. Then, the E/B decomposition of the $Q$ and $U$ maps 
was performed with this mask.  As for a full sky analysis, in order to
increase the signal 
to noise ratio, the E-mode maps were also
smoothed with a beam of FWHM=50'. Due to the apodization, most of the ambiguous modes are 
localized close to the mask edges. They can then be eliminated from the 
computation of the statistic by use of an additional mask applied to 
the derived E-mode map, but further broadened then apodized in a similar fashion to
the previous mask. In this case, apodization helps to minimize the
ringing effect which might appear on account of the use of the FFT
in computations of the matched circles statistic. In both
cases we used apodization with a Gaussian of FWHM=60'. Then,
the fraction of the unmasked part of the sky for the cut employed in
the computation of the statistic is $f_{\rm sky} = 0.55$. It is much less 
than for the original mask with the fraction $f_{\rm sky} = 0.73$. The masks are
shown in \fig\ref{fig:masks}. Such an aggressive excision of data
corresponds to a rather extreme case. We may hope that the broader frequency
range and lower noise level of the Planck data will enable us to
significantly extend the useful sky coverage for cosmological analysis.

This procedure for the computation of the E-mode signal does not remove completely
leakage from the B-modes. However, the residual is no
larger than the noise level in smoothed polarisation maps that is
negligible compare to the prevailing of E-mode signal. It is worth noticing that in
our procedure the residual signal from B-modes is additionally damped
by smoothing.

The false detection levels established from the requirement that fewer
than 1 in 100 simulations should yield a false match  are
shown in \fig\ref{fig:smax_t222_pol} corresponding to two 
extreme cases: analysis on the full sky and on a map with a
very conservative mask. The detection threshold for both
Planck and COrE should lie somewhere between these two
curves. However, more detailed studies of the influence of the Galactic
foreground polarisation is required, but is beyond the scope of this paper.

The minimum radius, $\alpha_{\rm min}$, of the correlated circles which can be detected for
a given map is determined by the intersection of the peaks in the 
matching statistic with the false detection level. The height of
the peaks indicates that the minimum radius is around $20^\circ$ for the Planck
and around $5^\circ$ for the COrE data in the case of the masked
map. If we do not detect any pairs of matched circles, this radius
will determine a lower bound on the size of the fundamental domain.

\subsection{Impact of reionisation on the matched circles statistic} \label{sec:reionisation}
One of the secondary anisotropy effects which might degrade our ability to
detect pairs of matched circles in polarisation induced by the
Thomson scattering of the CMB photons on the reionised matter. To evaluate
the level of impact of the effect on the performance our statistic, we
simulated the CMB map for the multi-connected universe both with and without
reionisation and compared the statistics.
The optical depth for the reionisation effect
was set as $\kappa=0.088$. This comparison showed that the relative
difference between the statistics is of order of $10^{-3}$. Therefore,
the influence of polarisation generated after reionisation is
negligible for the detectability of the matched circles. This should
not be a
surprising conclusion since the $S_{\rm
  max}$ statistic is sensitive rather to small angular scales while
the polarisation effect of reionisation prevails at large scales. In
case of the E-mode map the small scales are also additionally
amplified by the $\sqrt{(\ell+2)!/(\ell-2)!}$ factor, thus the large
angular scales are not so significant as for the temperature anisotropy.

\section{Conclusions} \label{sec:conclusions}
We have studied the possibility of using CMB polarisation maps for
constraining the topology of the Universe by means of the matched
circles statistic. We have shown that the CMB polarisation maps are free
from the degrading effect of the Doppler term that contributes to the
temperature anisotropy and that consequently allows matched
circles to be detected with a radius much smaller than for the CMB temperature
maps thus providing tighter constraints on the topology.
Moreover, the detection is robust with respect to the secondary
CMB polarisation generated after reionisation. However, the application
of polarisation data to studies of the topology of the Universe
depends to a large extent on systematic effects such as  
noise and Galactic foreground residuals.

Using simulations of the Planck polarisation data we demonstrated that
the level
of the instrumental noise should be sufficiently low  to enable the detection of
the signatures of a multi-connected topology, though it will not be
low enough to improve constraints on the topology derived from the
temperature maps. Nevertheless, it can serve as an important cross-check of the
latter. The best chance for the
detection of the matched circles is afforded by a map obtained from the inverse
noise weighted combination of the three least noisy Planck
frequency maps i.e.~100, 143 and 217 GHz maps. The polarised Galactic
emission should also be the weakest for the first two of these
channels enabling the best estimation of the CMB polarisation.

All the advantages of the polarisation analysis can be fully exploited
in the case
of the planned next generation of full sky low-noise
experiments. To demonstrate this,  we used simulations with
the resolution and sensitivity corresponding to the COrE mission. In
this case the smallest radius of the detectable back-to-back matched
circles in the polarisation maps is significantly smaller than 
for the temperature maps. An estimation based on our simulations shows
that the radius is around 5$^\circ$. 

The detectability of the matched circles depends heavily on our ability to
reliably separate  the CMB polarisation from the 
polarised emission of the Galaxy. We restricted our studies on this issue
only to the application of a mask intended to remove those areas of
the sky  most
contaminated by the Galactic foreground emission. We assumed
that the residuals of the Galactic emission outside the mask can be
sufficiently well modelled and subtracted from the map. However, it is
not certain to what extent this will be possible. As is well known,
removal of the polarised Galactic emission is a more challenging
problem than for the temperature maps since there is no
frequency band where the CMB polarisation dominates over
the polarisation of the Galactic foreground even for a small fraction
of the sky. 

We also have to bear in mind that a search for matched
circles on a masked map is able to constrain only those
universes with such dimensions and orientation of the fundamental domain
with respect to the mask that allow the detection of pairs of matched
circles. Since the probability
of overlooking circle pairs may be significant
for a mask that removes a large fraction of the sky, employing a cut as
small as possible is one of the crucial issues. The analysis of
a polarised version of an internal linear combination-type (ILC) map on the full sky
is also unlikely to be a credible solution of this problem. As shown in
\citet{bielewicz:2011} for the temperature case, Galactic foreground residuals
in the Galactic plane of an ILC map can cause significant spurious
correlations between the circles. In the case of polarisation maps it will
probably be an even more pronounced effect. However, it is to be hoped
that the high quality polarisation data from the Planck satellite will enable
a better understanding of the Galactic polarised emission and
therefore the use of a larger
fraction of the sky for cosmological analysis than for the \emph{WMAP} data.

\section*{Acknowledgments}
We acknowledge use of \textsc{camb}\footnote{http://camb.info/}
\citep{camb} and the \textsc{healpix} software \citep{gorski:2005} and
analysis package for deriving the results in this paper. This research
was supported by the Agence Nationale de la Recherche (ANR-08-CEXC-0002-01).


\end{document}